\documentclass[amsmath,preprintnumbers,9 pt,nofootinbib,amssymb,
 aps,twocolumn]{revtex4-2}
\usepackage{amsbsy}
\usepackage{amssymb}
\usepackage{graphicx}
\usepackage{color}
\usepackage{subfigure}
\usepackage{physics}
\usepackage{soul}
\usepackage{color}
\usepackage{bm}
\usepackage[normalem]{ulem}
\usepackage{bm}
\usepackage{array}
\usepackage{multirow}
\usepackage{lipsum}

\UseRawInputEncoding
\usepackage{verbatim}
\usepackage{natbib}
\begin{document}
\title{A mean field thermodynamic framework for time dependent self-assembly and pattern formation}
\author{Michael Nguyen$^{1,2}$, Suriyanarayanan Vaikuntanathan$^{1,2}$} 
\affiliation{$^1$The James Franck Institute, The University of Chicago, Chicago, IL,}
\affiliation{$^2$ Department of Chemistry, The University of Chicago, Chicago, IL.}
\begin{abstract}
In this work, we use a minimal model to introduce a framework for controlling self-assembly under the influence of time-dependent driving forces. We develop a mean-field thermodynamic framework that predicts the conditions required to reliably self-assemble a desired spatial pattern under time-varying external fields. We also calculate the entropy production associated with the time-dependent self-assembly process and examine how it can be used to predict conditions under which the external time-varying signal is reliably encoded as a spatial pattern in the self-assembling material. While the results in this work are developed in the context of a minimal one-dimensional model, we anticipate that the framework can be used to establish guidelines for controlling self-assembly in more complex scenarios. 

\end{abstract}
\maketitle

\section{Introduction}
Self-assembly is a process in which components in a system organize themselves into structures and patterns without human intervention. Though design principles for self-assembly for systems at global or local equilibrium have seen many advances recently, there has not been much progress for self-assembly far from equilibrium. Indeed, many essential processes
in biophysics and chemistry occur far from equilibrium~\cite{Cates2015,Battle604,Driscoll2017,Corte2008,Nguyen2014,Redner2013,Marchetti2013,Solon2014,Mann2009,Lan2012,Lele2013,Tu2008}. In our previous works~\cite{Nguyen2016,Nguyen2018,Nguyen2020}, using the ideas of stochastic thermodynamics we have developed a general framework to predict the structure of an assembly at steady state. Specifically, we can write bounds that relate the non-equilibrium driving forces to structures and basic kinetics of the assembly. Our framework, however, only applies to steady-state assemblies and does not extend to other important types of assemblies such as ones under time-periodic force.
Indeed, time-periodic forcing has been used to create non-equilibrium states with enhanced order in many contexts~\cite{SimonO.Lumsdon2004,Singh2009,Tagliazucchi2014,Corte2008,Helbing2000,Stanley2000,Evers2013,Nunes2016}. It was showed that dynamic phase transitions can occur by tuning the frequency of inputting the energy into the system~\cite{Klajn2009,CHENG2020}. In this
paper, we fill this gap by developing a general framework that can be used to explore how the pattern of the assembly forms under periodic drive and its connection to dissipation.


When the assembly is at a steady state with little to no internal relaxation, one can write out mean-field equations that relate the growth rates of particles with the components inside the assembly~\cite{Nguyen2016,Nguyen2018, gaspard2014kinetics}. Under the periodic drive, the self-assembled system is no longer in a time-independent steady-state, and the aforementioned mean-field conditions are no longer valid. The mathematical and thermodynamic treatment of such steady states is not trivial~\cite{Zheng2013}. As our first result, we show that many of our mean-field results~\cite{Nguyen2016,Nguyen2018} can still be adapted and used to probe the periodic assembly effectively under certain conditions. Further, inspired by works in~\cite{BaratoPRE2013,Barato2013MutualInfo,Verley2014,Hartich2014,Horowitz2015,BaratoClock2016,Horowitz2016}, we also develop a more general treatment that works for a broader set of conditions using bi-partite Markov Networks. Besides connecting the growth rates of the assembly with its inner structure, these models also allow us to write out the entropy production for the self-assembly process. The field of stochastic thermodynamics has shown how a tabulation of the entropy production rate can be used to efficiently constrain the non-equilibrium steady-state properties of a system. For example, the entropy production has been highlighted as measure of a system’s distance from equilibrium~\cite{Roldan2010,Fodor2016,Seara2018,Li2019,Frishman2020,Martinez2019} and has been used to characterize critical behavior in non-equilibrium phase transitions~\cite{Gaspard2004,Tom2012,Noa2019,Zhang2016}. In our previous works~\cite{Nguyen2016,Nguyen2018,Nguyen2020}, we had demonstrated that the entropy production can be used to predict the structure of non-equilibrium assembly. Following these ideas, in this paper, we show that the entropy production informs how time correlation between particles affect the patterns being deposited into the assembly.


The paper is organized in the following way. In Section~\ref{sec:model}, we introduce the model and simulations used in this paper and explore their properties. In Section~\ref{sec:meanfield}, we review the mean-field approach that was developed in~\cite{Nguyen2014,Nguyen2018,gaspard2014kinetics} for steady-state setting and demonstrate how it is still applicable for a periodic system under certain regimes. We then develop the bi-partite Markov model in Section~\ref{sec:bipartitle} which is applicable in a broader range of conditions and can be used to predict the properties of time-dependent non-equilibrium steady states with surprising accuracy. Finally, we look at the entropy production of the two models in Section~\ref{sec:entropy} and show how they can be used to investigate how patterns are driven into the assembly.

\section{Model And Simulations
\label{sec:model}}
\begin{figure}
    \includegraphics[width=0.66\linewidth,scale=0.3,angle=270]{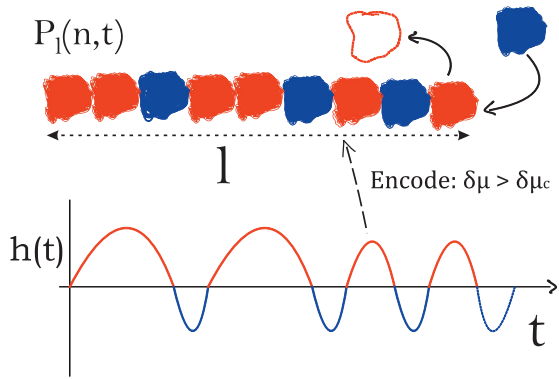}
    \caption{Schematic of the one dimension growth assembly with rates affected by a time-varying external field $h(t)$. A polymer is grown from a bath containing two distinct monomers labelled by $R$ and $B$. The rates of addition and removal are influenced by the energy of interaction between the various monomers, an  oscillating external field, and the chemical potential of the monomers in the bath. In this work, we show that a minimal extra chemical potential from equilibrium, $\delta\mu_c$, is required to consistently encode the signal of the external field into the assembly. We also outline a mean field theory to estimate the probability of obtaining a monomer of type $n$ ($R/B$ in this case) a fixed distance $l\gg l_m$, where $l_m$ is the length scale associated with a single monomer, away from the tip of the assembly.}
    \label{fig:timedependentsystem}
\end{figure}

We consider a one-dimensional growth of an assembly made up of two kinds of particles (denoted by $R$ and $B$) as shown in Fig.~\ref{fig:timedependentsystem}. In order to compactly specify the energetics of this system, we will assign numeric values of  $1$ and $-1$ to the two monomer types $R$ and $B$, respectively. As the assembly grows, we only allow particles to attach or detach at the outermost block of the assembly. We choose dynamics wherein particles are added into the assembly at a constant rate:
\begin{equation}
\begin{split}
    k^{add}_t({n_i,n_j\xrightarrow{}n_j,n_k})&=\exp(\mu)P_{gen}(n_k|n_j)\\
    \label{eq:additionrate}
\end{split}
\end{equation}
Here n represents the configurations/properties of the particles that make up the assembly (in this case $1$ or $-1$). $P_{gen}(n_k|n_j)$ is the probability of getting a particular  particle $n_k$ from the bath given the interface particle is $n_j$. This probability depends on the relative concentrations of the particles in the bath. For this paper, we will let the two particles have equal concentrations setting $P_{gen}(n_k|n_j)=0.5$. $\mu$ is the chemical potential of the particles on the bath.
The particles, however, are removed with a time-dependent rate:
\begin{equation}
    k^{rem}_t(n_k,n_i\xleftarrow{}n_i,n_j)=\exp(-Jn_{i}n_{j}-h\cos{\left(\frac{2\pi t}{T}\right)}n_{j})
\end{equation}
Here $J$ specifies the monomer-monomer interaction strength, $h$ is the amplitude of the external field, $t$ is time, and $T$ is the period.  When there is no external field, $h=0$, the assembly will reach a steady-state, and its configuration will be set by the chemical potential $\mu$. Specifically, when $\mu=\mu_{\rm{eq}}$, the assembly is at equilibrium and does not grow on average. When $\mu>\mu_{eq}$, the assembly grows, while for $\mu<\mu_{eq}$, it will shrink. Throughout this paper, we will write $\mu = \delta\mu + \mu_{\rm{eq}}$ with $\delta\mu>0$ to denote the extra chemical driving force on the growing assembly. We then run simulations with the above rates using the First Reaction Method (FRM) \cite{JANSEN1995}. 

\begin{figure}
    \centering
    \includegraphics[width=1\linewidth,angle=0,scale=0.9]{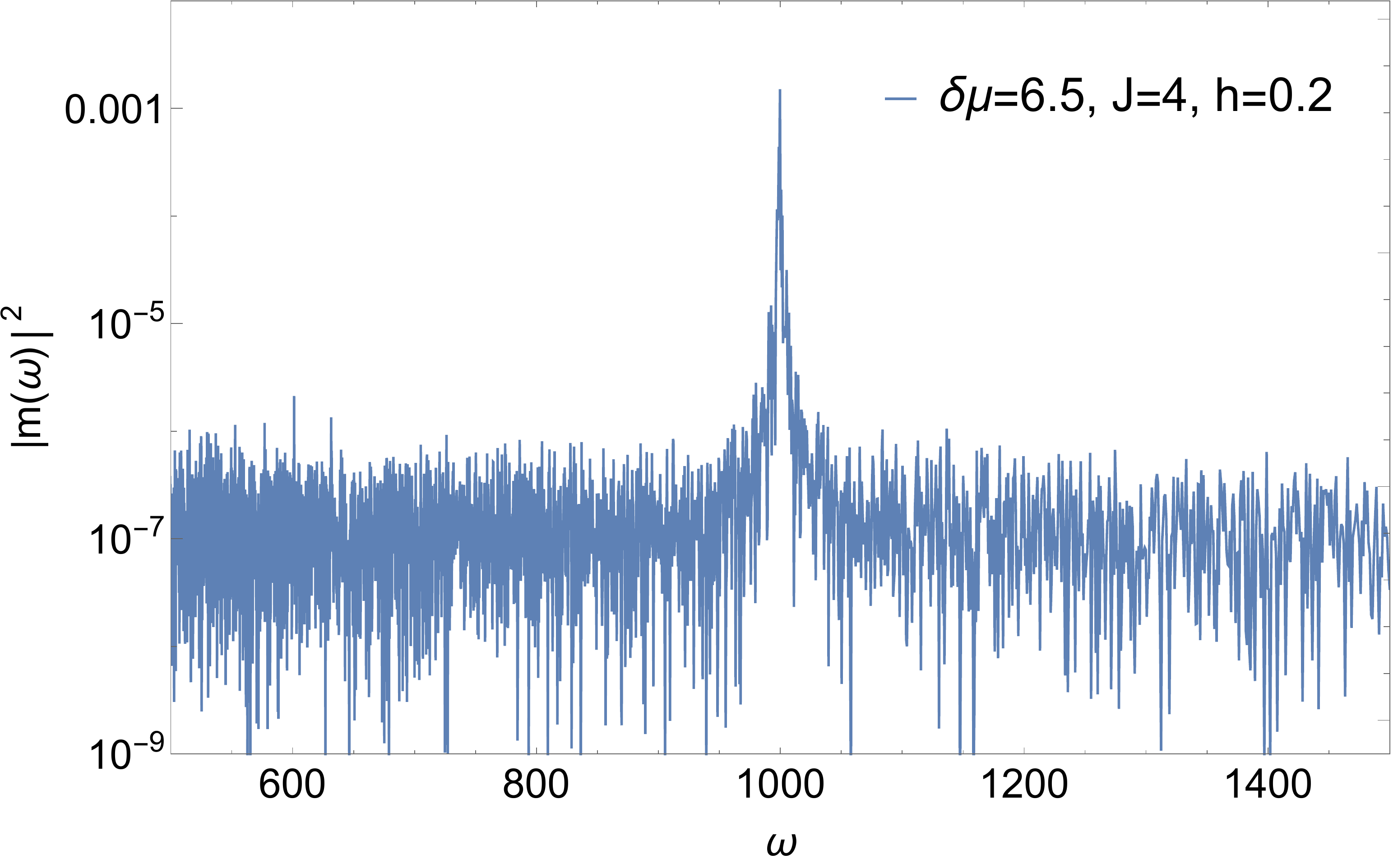}
    \caption{Fourier Transform of the assembly obtained using simulations at $\delta\mu=6.5$, $J=4$, $h=0.2$ and $T=1000$. The period of the peak is the same as the period of the external field.}
    \label{fig:FourierTimeDep}
\end{figure}

\begin{figure}
    \centering
    \includegraphics[width=1\linewidth,angle=0,scale=0.9]{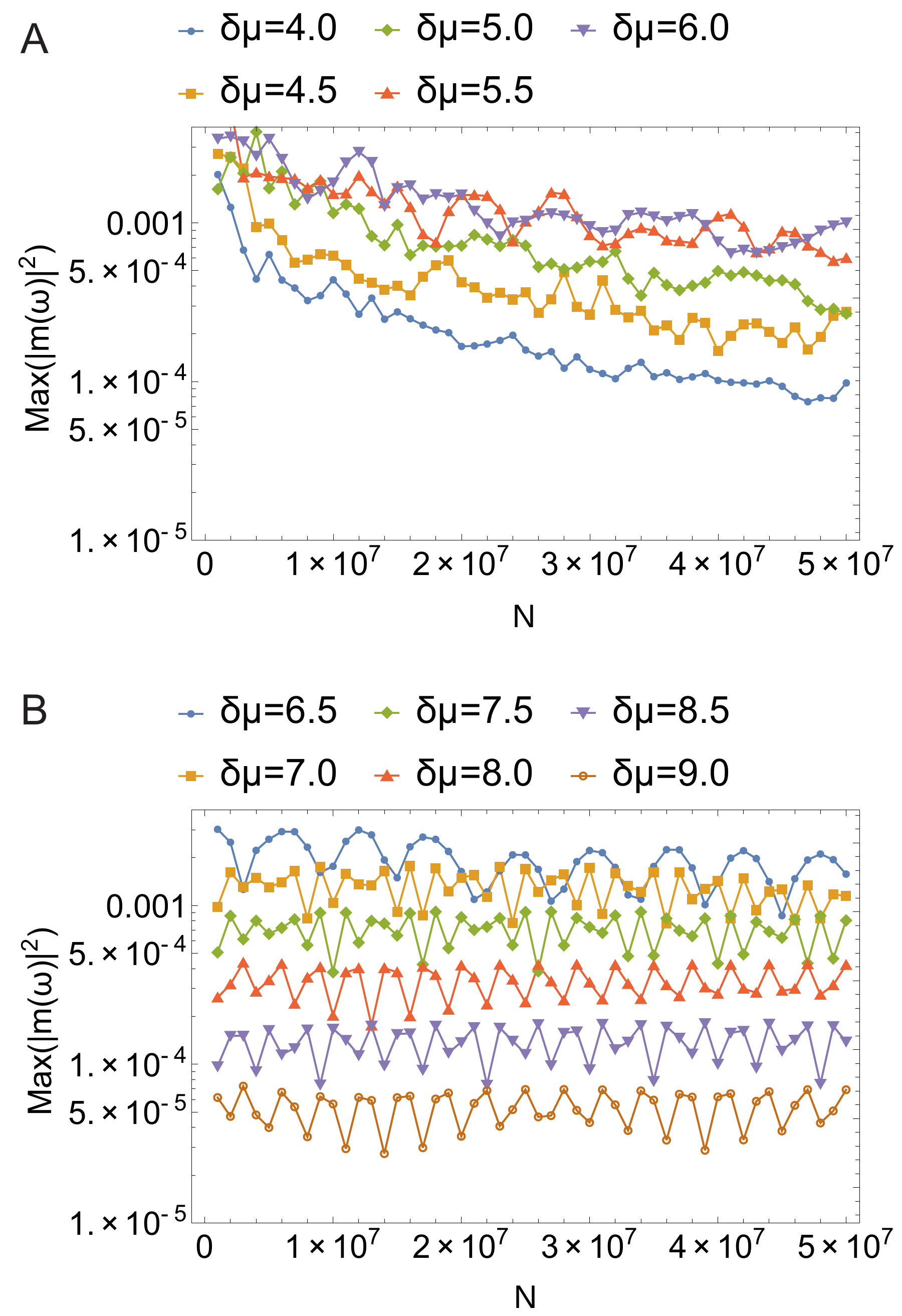}
    \caption{Peak of the Fourier Transform as a function of the number of particles in the assembly with parameters: $J=4$, $h=0.2$ and $T=1000$. Top: in this range, $\delta\mu = 4.0$ to $6.0$ , the peak keeps decaying as the number of particles increases. Down: in this range, $\delta\mu = 6.5$ to $9.0$, the peak stays constant with the number of particles.}
    \label{fig:FourierProfile}
\end{figure}

\begin{figure}
    \centering
    \includegraphics[width=1\linewidth,angle=0,scale=0.9]{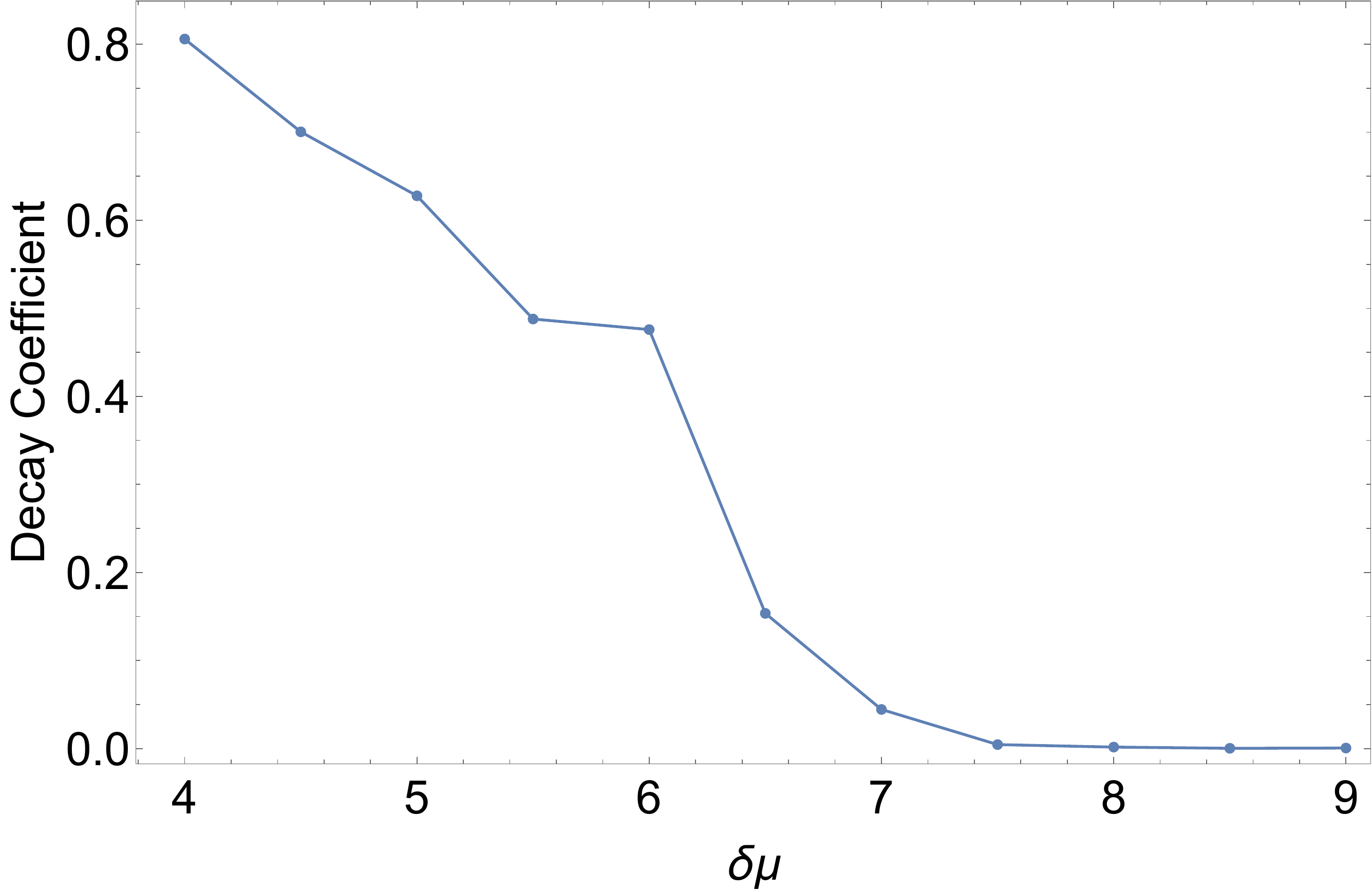}
    \caption{Decay coefficient of the assembly as a function of 
    $\delta\mu$ at $J=4$, $h=0.2$ and $T=1000$.. The decay coefficients, a, are extracted using $\ln{\max(|m|^2)} = a\ln{N}+b$.  }
    \label{fig:DecayCoefs}
\end{figure}

To analyze how well the external field $h(t)$ affects the pattern inside the assembly, we take the whole assembly generated by the simulation through a Fourier Transform with the convention:
\begin{equation}
    m(n)=\frac{1}{N}\sum _q^N  m(q) \exp(\frac{2\pi i qn}{N})
\end{equation}
Here $m(n)$ denotes the state of the particle at a certain position $n$, $N$ is the total number of particles in the assembly, and $q$ is the spatial frequency of the assembly's pattern. We then convert the spatial frequency into the time period, $\omega$, using the growth rate: $ \dot{N}_{\rm{ave}}=\frac{1}{T}\int_0^T\dot{N}(t)dt=\frac{1}{q\omega}$. Fig.~\ref{fig:FourierTimeDep} shows the result of the Fourier Transform for an assembly consisting of $10^7$ particles. The location of the peak of the Fourier Transform corresponds to the period, $T$, of the external field. We repeat this analysis to assemblies obtained with $\delta\mu$ from the range: $\delta\mu: 4.0 - 9.0$ and plot the magnitude square of the peak varying with the number of particles in the assemblies in Fig.~\ref{fig:FourierProfile}.
From Fig.~\ref{fig:FourierProfile}, we see that the peak of the assembly first increases with increasing $\delta\mu$. It then caps out and decreases with increasing $\delta\mu$.  Closer inspection also reveals that the peaks decay with an increasing number of particles.
To quantify the decay, we fit the peak to the number of particles using the formula: $\ln{\max(|m|^2)} = a\ln{N}+b $. The parameter $a$ is the decay coefficient of the assembly peak. Fig.~\ref{fig:DecayCoefs} shows how this decay coefficient changes with $\delta\mu$:  With $\delta\mu < 7$, the decay coefficient is positive, and the pattern in the assembly is not deposited consistently throughout the assembly. Only with $\delta\mu > 7$, the pattern is deposited consistently into the assembly in every cycle, which results in a $0$ decay coefficient. This suggests that to effectively drive a pattern from a varying field into the assembly, a minimum amount of driving force, $\delta\mu$, is required.



\begin{figure}
    \centering
    \includegraphics[width=1\linewidth,angle=0,scale=0.9]{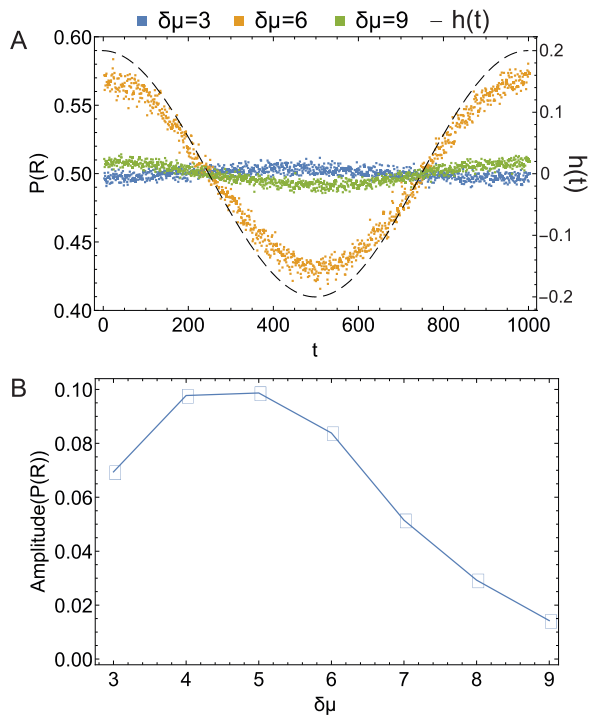}
    \caption{Up: Probability profiles of obtaining a red particle as a function of time. The profile of the external field, $h(t)$, is also plotted in dash line for reference. Down: Amplitudes of the probability profiles of obtaining a red particle. These results are obtained from simulations with $J=4$, $h=0.2$ and $T=1000$.}
    \label{fig:ProbabilityProfileInside}
\end{figure}


In addition to the correlations obtained from the Fourier Transform, another figure of merit to ascertain how the correlations inside the assembly are guided by the time-dependent external field can be constructed by measuring the probability to find a specific monomer type in the bulk of the assembly, a fixed distance away from the tip of the assembly~\ref{fig:timedependentsystem}, $P^t_{inside}(R/B)$. Mirroring the time dependence of the external field, this probability function will also be a function of time. 
In practice, we extract this probability by recording the moment in which a particle at the interface is covered by a new particle coming into the assembly. For example, after growing at least $5\cross 10^7$ particles, the probability of obtaining a $R$ particle at a certain time is computed using:
 \begin{equation}
     P^t_{inside}(R)=\frac{N_R(t\%T)}{N_R(t\%T) +N_B(t\%T)}
 \end{equation}
 Here $t$ is the time in which the particle is added into the assembly, and $T$ is the period of the external field. Note that $P^t_{inside}(R/B)$ is distinct from $P^t(R/B)$, the probability associated with finding a monomer of a specific sort at the tip of the assembly.
 Fig~\ref{fig:ProbabilityProfileInside}A shows the probabilities inside the assembly collected at different $\delta\mu$.
Similar to the peaks of the Fourier Transforms, the amplitude of the probability profiles also maximizes at certain $\delta\mu$ as shown in Fig~\ref{fig:ProbabilityProfileInside}B.
 Note, however, the location of this peak is at $\delta\mu=5$ while the one obtained from the Fourier Transform is at $\delta\mu=7$. This offset is simply due to the fact that the probabilities  $P^t_{inside}(R/B)$ are single-particle quantities while the Fourier transform accounts for many-particle correlations. 
In the next section, we develop a mean-field theory for predicting the probability profiles. 
\section{Mean-Field Approach\label{sec:meanfield}}
We first review the mean-field approach that we develop in our previous works~\cite{Nguyen2016,Nguyen2018} for steady-state systems in the absence of time-varying fields.
Briefly, instead of looking at the full configurations of the assembly, we focus on the interface of the assembly. Specifically, the system of interest will consist of a particle that is in contact with the bath and the one right next to it. We then write out master equations for the interface probability: $P^{ss}(n_i,n_j)$ as shown in the appendix Eq.~\ref{eq:timeindependentequation}. The superscript $ss$ is there to indicate that the system is now at a steady state because the external field is no longer changing with time. For this particular 1D system, the rates for the system
are:
\begin{equation}
        k^{add}({n_i,n_j\xrightarrow{}n_j,n_k})=\exp(\mu_{eq}+\delta\mu)P_{gen}(n_k|n_j)
\end{equation}
\begin{equation}
        k^{rem}(n_k,n_i\xleftarrow{}n_i,n_j)=\exp(-Jn_{i}n_{j}-hn_{j})P_{sys}(n_k|n_i)
\end{equation}
Notice that the external field in the removal rate is no longer time-dependent. In addition, a new term, $P_{sys}(n_k|n_i)$ has appeared in the removal rate. This is the conditional probability of obtaining a particle $n_k$ in the assembly given its neighbor $n_i$. This probability arises from the uncertainty of not looking at the assembly's full configuration. We now make the mean-field assumption that the conditional probability, $P_{sys}(n_k|n_i)$, can be described using the bulk structure of the assembly:
\begin{equation}
    P_{sys}(n_k|n_i)=\frac{P^{ss}_{inside}(n_k,n_i)}{\sum_{n_k}P^{ss}_{inside}(n_k,n_i)}
    \label{eq:conditionalprop}
\end{equation}
Here $P^{ss}_{inside}(n_k,n_i)$ is the probability of observing a configuration $(n_k,n_i)$ inside the assembly at steady state. To solve the master equations at steady state, we still need a relation between the probabilities inside the assembly, $P^{ss}_{inside}(n_k,n_i)$, and the probabilities at the interface $P^{ss}(n_k,n_i)$. This is established by noticing that the rate in which a particle is deposited into the assembly is proportional to its amount inside the assembly: 

\begin{equation}
    P^{ss}_{inside}(n_i)=\frac{\sum_{n_k,n_j}J^{ss}(n_k,n_i\rightarrow n_i,n_j)}{\sum_{n_k,n_i,n_j}J^{ss}(n_k,n_i\rightarrow n_i,n_j)}
    \label{eq:naivemeanfield}
\end{equation}
Here $J^{ss}(n_k,n_i\rightarrow n_i,n_j)$, the current as the assembly grows from from $(n_i,n_j)$ into $(n_j,n_k)$, is defined as:

\begin{equation}
\begin{split}
  &J({n_i,n_j\xrightarrow{}n_j,n_k})=\\
  &P^{ss}(n_i,n_j)P_{gen}(n_{k}|n_{j})k^{add}({n_{i},n_{j}\xrightarrow{}n_j,n_{k}})\\
  &-P^{ss}(n_j,n_k)P_{sys}(n_i|n_j)k^{rem}({n_{i},n_{j}\xleftarrow{}n_j,n_{k}})
  \label{eq:currentdef}
\end{split}
\end{equation}
We can now solve the master equations Eq.~\ref{eq:timeindependentequation} at steady state by setting its left-hand side to zero and obtain the probability profiles inside the assembly.

{ As we vary the external field with time, the rates $k^{add}$ and $k^{rem}$ are in general time-dependent, and the system is no longer in a steady state. One way to extend the above mean-field approach is assuming
the system is in a quasi-steady state. In other words, in solving the master equation, the left-hand side of Eq.~\ref{eq:timeindependentequation} is still set to zero, while the rates on the right-hand side will correspond to a different point in time.} 
However, this naive approach only works in the limit where the oscillation is extremely slow (high $T$) or extremely high (small $T$).
In general, we cannot assume the system is at a quasi-steady state. We can, however, solve the master equations Eq.~\ref{eq:timeindependentequation}
numerically with time-dependent rates and use Eq.~\ref{eq:naivemeanfield} to predict the probabilities inside the assembly.
\begin{figure}
    \centering
    \includegraphics[width=1\linewidth,angle=0,scale=0.9]{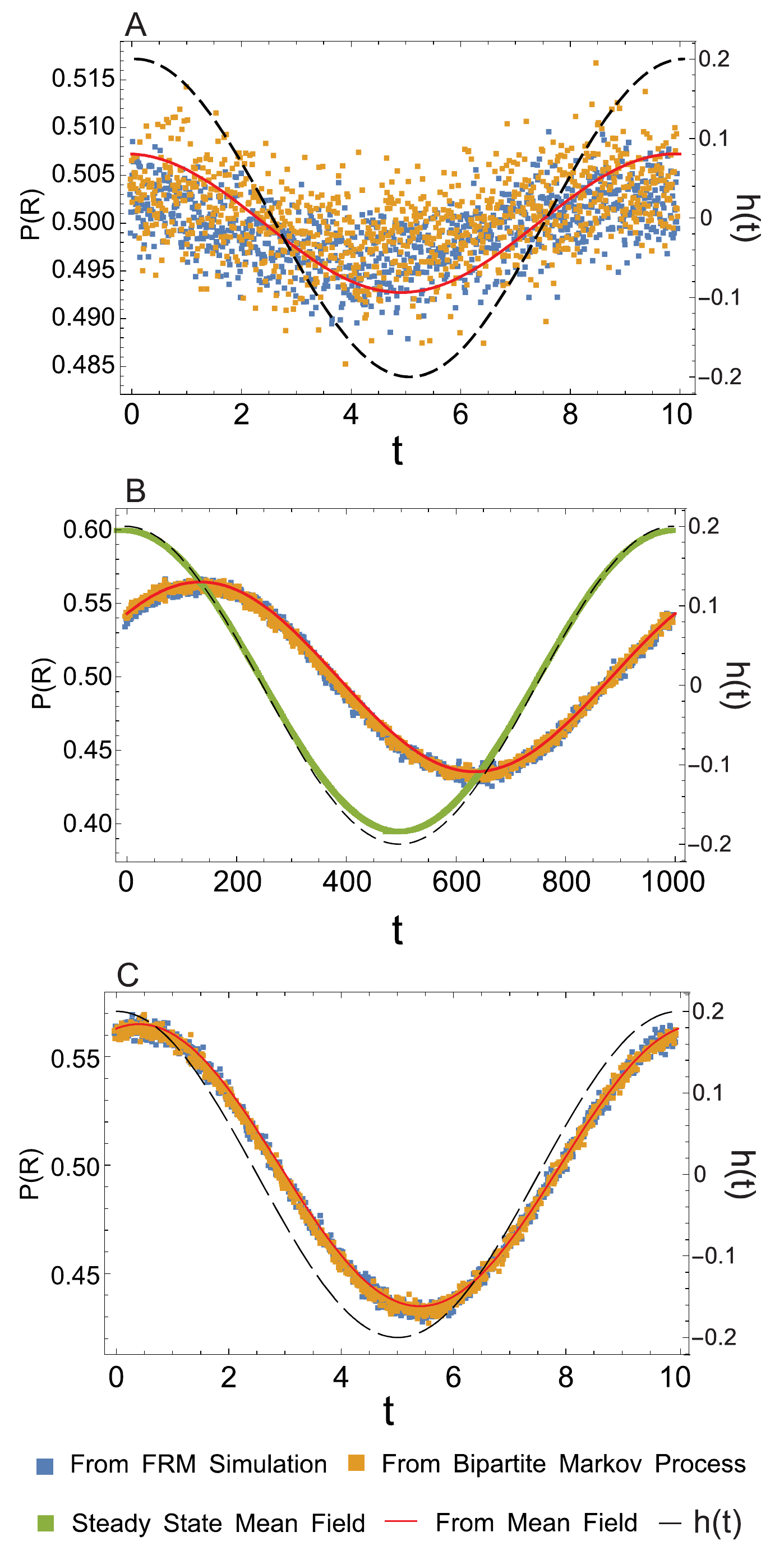}
    \caption{Probability profile for obtaining a red particle at a certain time in the assembly at A: This profile is at $\delta\mu=3, J=4, h=0.2, T=10$. B: This profile is at $\delta\mu=3, J=4, h=0.2, T=1000$. C: This profile is at $\delta\mu=6, J=4, h=0.2, T=10$. The blue dots are obtained from FRM simulations. The orange dots are obtained from simulations of the bi-partite Markov process with Eq.~S3 and S4. The red curve is obtained by solving the mean-field master equations, Eq.~\ref{eq:timeindependentequation}, numerically with time-dependent rates. The green dots are obtained from solving the mean-field master equations, Eq.~\ref{eq:timeindependentequation}, by assuming the assembly is at quasi steady-state at every point in time. }
    \label{fig:PropInsideAmpmu3omega10}
\end{figure}
In Fig~\ref{fig:PropInsideAmpmu3omega10}, we compare the probability profiles obtained from the FRM simulations (blue dots) and those from solving the mean-field master equation, Eq.~\ref{eq:timeindependentequation}, numerically (red curve), which indicates that in the regime of high $\delta\mu$ and high $T$, the mean-field master equations can be used to predict the structures inside a periodic assembly. This is because, in these regimes, the rates at which particles going into the assembly are relatively high compared to the rate that the field is changing. However, this does not mean the assembly is in a quasi-steady state as in the naive mean-field approach. In Fig.~\ref{fig:PropInsideAmpmu3omega10}B, we also plot the probability profile by solving the master equations using the naive mean-field approach (green dot) to contrast the one obtained from solving the time-dependent master equations numerically (red curve). 

In this section, we have shown that the mean-field framework developed in~\cite{Nguyen2016,Nguyen2018,gaspard2014kinetics} for a steady-state assembly can be extended to a periodic assembly. With this approach, instead of focusing on the full configuration of the assembly, we can look at the kinetics in the interface then deduce the assembly's configuration. The predictions from the mean-field are particularly accurate in the regime of high $\delta\mu$ and high $T$ where the kinetics of the assembly is relatively fast to the oscillating field.
\section{Bi-Partite Markov Network \label{sec:bipartitle}}
When we develop our mean-field model, one assumption we take is that the conditional probability, $P_{sys}(n_k|n_i)$, is related to the probability of the bulk $P^{ss}_{inside}(n_k,n_i)$ through Eq.~\ref{eq:conditionalprop}. This only applies when the kinetics of the assembly is fast relative to the field.
When the kinetics of assembly is comparable to the oscillation of the field, however, different parts of the assembly will experience different parts of the field, and we cannot use Eq.~\ref{eq:conditionalprop} to predict the particle $n_k$ given just its neighbor $n_i$. 
To take in account this effect, we have to include a time-dependent term in our interface probabilities: $P^t(n_{i}t_x,n_{j}t_y)$. Here in the notation $n_{i}t_x$, the $t_x$ represents the time that the particle $n_i$ is added into the assembly. In order words, a particle will remember the time it is deposited into the assembly. The $t_y$ on the interface particle, $n_j$, represents the current time of the system.
\begin{figure}
    \centering
    \includegraphics[width=1\linewidth,angle=270,scale=0.7]{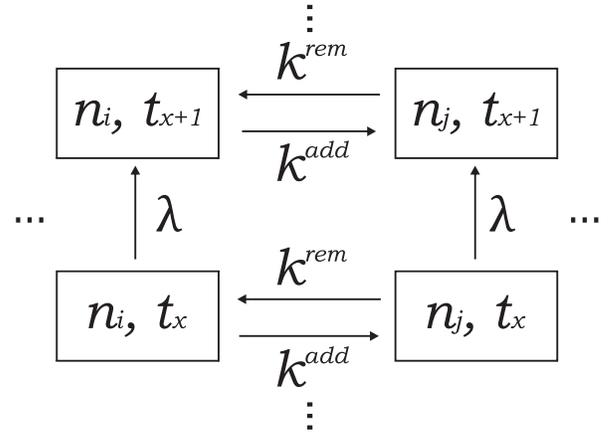}
    \caption{Schematic of the bi-partite Markov model. In addition to addition and removal of particles (horizontal transitions), the system now has moves that advance time, $t_x$, at the interface (vertical transitions).}
    \label{fig:BiPartiteMarkov}
\end{figure}
 To develop an improved mean field theory that accounts for such effects, inspired by works in~\cite{BaratoPRE2013,Barato2013MutualInfo,Verley2014,Hartich2014,Horowitz2015,BaratoClock2016,Horowitz2016}, we set up a bi-partite Markov network between time $t$ and configuration of the assembly $n$. As shown in Fig.~\ref{fig:BiPartiteMarkov}, besides the addition and removal of particles, the system now has a move that advance time $t_y$ into $t_{y+1}$. The new move is constructed as following: First, the period $T$ is partitioned into $M$ parts. The time $t_y$ then moves across these partitions with rate $\lambda=\frac{M}{T}$ as demonstrated in Ref.~\cite{BaratoClock2016}. When the system advances in $t_y$, the external field also changes according to $h\cos{(\frac{2\pi t_y}{T})}$ which in turn affects the removal rate $k^{rem}$. With this setting, all the possible transitions in this bi-partite Markov process can be summarized as follow:

\begin{equation}
    \begin{split}
        &{\bf (1)}\quad n_{i}t_x,n_{j}t_y\genfrac{}{}{0pt}{2}{\xrightarrow{ k^{add}}}{\xleftarrow[ k^{rem}]{}}n_{j}t_y,n_{k}t_y\\
        &\\
        &{\bf (2)}\quad n_{i}t_x,n_{j}t_y{\xleftarrow[k^{rem}]{}}n_{j}t_z,n_{k}t_y\\
        &  (\text{with}\quad t_z \neq t_y)\\
        &\\
        &{\bf (3)}\quad n_{i}t_x,n_{j}t_y\quad{\xrightarrow{\lambda}}\quad  n_{i}t_x,n_{j}t_{y+1}
    \label{eq:addremoveparticletrans}
    \end{split}
\end{equation}

\begin{figure}
    \centering
    \includegraphics[width=1\linewidth,angle=0,scale=1]{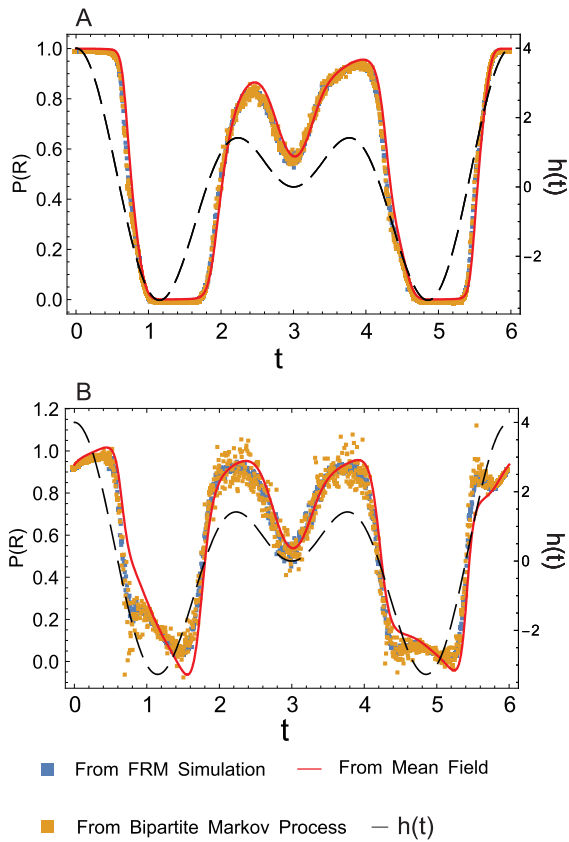}
    \caption{Probability profile for obtaining a red particle at a certain time in the assembly with two external fields at different periods $T_1=2$ and $T_2=3$ with $J=4, h=2$. A: This profile is for $\delta\mu=5$. B: This profile is for $\delta\mu=3$. The results from bi-partite Markov processes agree very well with ones from FRM simulations. The mean-field method, however, only works well at high $\delta\mu$. }
    \label{fig:PropInsideAmpmu5harmonic2}
\end{figure}

We can now write the master equations for the bi-partite Markov process as shown in the appendix (Eq.~S2 and Eq.~S3). After solving the bi-partite Markov master equations consistently at a steady-state, we can connect the probabilities at the interface and the probabilities inside the assembly using the currents as we did in Eq.~\ref{eq:naivemeanfield}:
\begin{equation}
\begin{split}
    &P^{t}_{inside}(n_it_x)\\
    &\propto\sum_{n_j,n_k,t_y,t_z}\Big[J^{t}_{add}({n_{k}t_y,n_{i}t_x\xrightarrow{}n_it_x,n_jt_x})\\
    &-J^{t}_{rem}({n_{k}t_y,n_{i}t_z\xleftarrow{}n_it_x,n_jt_z})\Big]
    \label{eq:firstmeanfieldequation}
\end{split}
\end{equation}
With
\begin{equation}
    \begin{split}
       &J^{t}_{add}({n_{k}t_y,n_{i}t_x\xrightarrow{}n_it_x,n_jt_x})=\\
    &P_{gen}(n_j|n_i)k_t^{add}({n_{k}t_y,n_{i}t_x\xrightarrow{}n_it_x,n_jt_x})P^{t}(n_kt_y,n_it_x) \label{eq:currenttimeadd}
    \end{split}
\end{equation}
and
\begin{equation}
    \begin{split}
       &J^{t}_{rem}(n_kt_y,n_it_z\xleftarrow{}n_{i}t_x,n_{j}t_z)=\\
    &P^{t}(n_it_x,n_jt_{z})P^t_{sys}(n_kt_y|n_it_z)k_t^{rem}(n_kt_y,n_it_z\xleftarrow{}n_{i}t_x,n_{j}t_z)\ \label{eq:currenttimerem}
    \end{split}
\end{equation}
Figs.~\ref{fig:PropInsideAmpmu3omega10} show that the probability profiles obtained using Eq.~\ref{eq:firstmeanfieldequation} (orange dots) match very well with ones from FRM simulations (blue dots). We have also tested Eq.~\ref{eq:firstmeanfieldequation} using different kinetics and harmonics; Eq.~\ref{eq:firstmeanfieldequation} (orange dots) still captures the probability profiles very well when the mean field method (red curve) does not (Fig.~\ref{fig:PropInsideAmpmu5harmonic2}). 

\section{Entropy Production \label{sec:entropy}}
In the previous sections, we have shown that two approaches, mean-field and bi-partite Markov process, do an excellent job in reproducing the probability profiles of the periodic assembly in their proper regimes. These results also clarified conditions under which the temporal oscillations in the external signal could be encoded in the assembly. We now explore whether an accounting for the entropy production rate can also provide intuition for this problem. 
\subsection{Entropy Production From the Time Dependent Mean Field Approach}
In our previous works~\cite{Nguyen2016,Nguyen2018}, we have written down a general form for the entropy production of a self-assembly process with the mean-field approach:
\begin{equation}
    \dot{S}_{\rm{mf}}=\dot{N}(\delta\mu - \epsilon_{diss})
    \label{eq:entropyprodsteady}
\end{equation}
where $\epsilon_{diss} =\langle E^{\rm{eq}}\rangle - G^{\rm{eq}} - \langle E^{\rm{eff}}\rangle + G^{\rm{eff}} $.
Here $E^{\rm{eq}}$ is the equilibrium energy for our system. 
$E^{\rm{eff}}$ is the effective energy that corresponds to structure of the system when it is out equilibrium. 
The bracket $\langle .. \rangle$ is the ensemble average of the system's non-equilibrium configuration. $G^{\rm{eq}}$ is the free energy corresponding to the equilibrium energy while $G^{\rm{eq}}$ is the free energy corresponding to the effective energy. Eq.~\ref{eq:entropyprodsteady} shows that the entropy production rate doesn't depend on the detailed kinetics of the process but is simply captured in terms of the difference between the non-equilibrium and equilibrium steady states and an account of the non-equilibrium driving forces.

In Section~\ref{sec:meanfield}, we have extended the mean-field approach to an assembly under period drive. Along those lines with the formalism of time-dependent entropy production \cite{Esposito2010}, we can also readily adopt the form of the entropy production for the steady-state assembly, Eq.~\ref{eq:entropyprodsteady}, to the periodic assembly:
\begin{equation}
    \dot{S}_{\rm{mf}}=\frac{1}{T}\int_0^T \dot{N}(t)(\delta\mu(t) - \epsilon_{diss}(t))dt
    \label{eq:entropyprodmeanfield}
\end{equation}
where $\epsilon_{diss}(t) =\langle E^{\rm{eq}}\rangle(t) - G^{\rm{eq}}(t) - \langle E^{\rm{eff}}\rangle(t) + G^{\rm{eff}} (t)$.

For our system, the $E^{\rm{eq}}(t)$ is the energy of the 1D polymer model with coupling constant $J$ and a constant external field $h$:
\begin{equation}
    E^{\rm{eq}}(t)=\sum_{i}-Jn_in_{i+1}-h\cos{\left(\frac{2\pi t}{T}\right)}n_{i}
\end{equation}
For the $E^{\rm{eff}}(t)$, we use an effective coupling constant $J_{\rm{eff}}$ and effective external field $h_{\rm{eff}}$ to describe the configurations in the assembly at a point in time:
\begin{equation}
    E^{\rm{eff}}(t)=\sum_{i}-J_{\rm{eff}}n_in_{i+1}-h_{\rm{eff}}\cos{\left(\frac{2\pi t}{T}\right)}n_{i}
\end{equation}
The bracket $\langle .. \rangle$ now is the ensemble average of the system at equilibrium with the energy $E^{\rm{eff}}(t)$. The effective coupling constant $J_{\rm{eff}}$ and effective external field $h_{\rm{eff}}$ can be extracted using observables. Specifically, for our system, we can use results of the 1D-Ising model to relate $\langle n_i \rangle (t)$, $\langle n_in_{i+1} \rangle (t)$ with $J_{\rm{eff}}$ and $h_{\rm{eff}}$ as demonstrated in \cite{Nguyen2016}. The $\delta\mu(t)$ is computed using $\delta\mu(t)=\mu-G^{\rm{eq}}(t)$. We plot the entropy production from Eq.~\ref{eq:entropyprodmeanfield} in Fig.~\ref{fig:EntropyProductionTimeDep}.

As the system approaches the $\delta\mu$ in which pattern is added into the assembly consistently (around $\delta\mu\approx 7$), this entropy production rate approaches an inflection point. The inflection point is due to the growing contributions from the $\epsilon_{\rm diss}$ term. Indeed, the quantity $\epsilon_{\rm diss}$ can also be written as a relative entropy between the observed distribution of patterns in the assembly and the corresponding equilibrium steady-state distribution. As periodic spatial correlations begin to get reliably encoded in the assembly, the increase in the relative entropy captured by $\epsilon_{\rm diss}$ can be sufficient to counteract $\delta \mu$. This can lead to saturation or slow down in the rate of increase of entropy production. 

These results suggest that an analysis of the entropy production rate may be used to predict the minimum non-equilibrium forces required to reliably sustain correlations in the non-equilibrium assembly.
\subsection{Entropy Production From Bi-Partite Markov Process}
\begin{figure}
    \centering
    \includegraphics[width=1\linewidth,angle=270,scale=0.7]{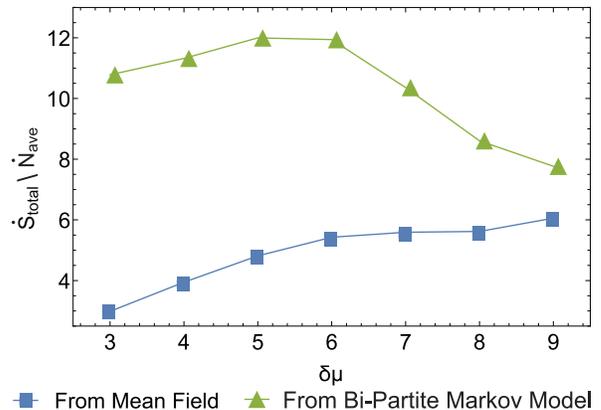}
    \caption{Entropy production of the one dimension growth assembly with rates affected by a time-varying external field h(t) with $J = 4, h = 0.2, T = 1000$. $\dot{N}_{\rm{ave}}$ is the average growth rate of assemblies over the period T: $\dot{N}_{\rm{ave}}=\frac{1}{T}\int_0^T\dot{N}(t)dt$. The blue curve is computed using the mean-field entropy production, Eq.~\ref{eq:entropyprodmeanfield}. The green curve is computed using the full bi-partite Markov Process entropy production, Eq.~\ref{eq:totalentropyproductionplusobserve}.
    }
    \label{fig:EntropyProductionTimeDep}
\end{figure}

Similar inferences can be drawn by studying the entropy production rate of the bi-partite Markov state caricature. 
Because of the setup of the bi-partite Markov process, some of the transitions are unidirectional, so the normal definition of entropy production does not apply to each transition. However, according to \cite{Busiello2020}, the total entropy production of the system does not depend on these unidirectional transitions but only on the bidirectional transitions \cite{Busiello2020}. The total entropy production, thus, is:
\begin{equation}
\begin{split}
    &\dot{S}=\sum_{t_x,t_y,n_k,n_i,n_j}\big[J^{t}_{add}({n_{k}t_y,n_{i}t_x\xrightarrow{}n_it_x,n_jt_x})\\
    &-J^{t}_{rem}(n_kt_y,n_it_x\xleftarrow{}n_{i}t_x,n_{j}t_x)]\\
    &\ln{\frac{J^{t}_{add}({n_{k}t_y,n_{i}t_x\xrightarrow{}n_it_x,n_jt_x})}{J^{t}_{rem}(n_kt_y,n_it_x\xleftarrow{}n_{i}t_x,n_{j}t_x)}}
    \label{eq:totalentropyproductionplusobserve}
\end{split}
\end{equation}
Here the currents $J^t_{add}$ and $J^t_{rem}$ are defined according to Eq.~\ref{eq:currenttimeadd} and Eq.~\ref{eq:currenttimerem}.  Thus, after solving the master equations of the bi-partite Markov process (Eq.~S2 and Eq.~S3) at steady state, Eq.~\ref{eq:totalentropyproductionplusobserve} can be used to calculate the entropy production of the process. In Fig.~\ref{fig:EntropyProductionTimeDep}, we plot the entropy production of the bi-partite Markov process (Eq.~\ref{eq:totalentropyproductionplusobserve} -green triangle) with the entropy production of the mean-field equations (Eq.~\ref{eq:entropyprodmeanfield}-blue square). This shows that the entropy production of the bi-partite Markov process is always higher than the one of the mean-field equations. This is because we have artificially labeled timestamps into the particles. Reflecting the increased information content, the entropy production also increases. 
As shown in Sec.~\ref{sec:meanfield}, in the limit of high $\delta\mu$ and $T$, the bipartite master equation approaches the mean-field master equations derived previously, and the two entropy production rates converge in this limit. Reflecting these limits, the entropy production rate of the bi-partite Markov process has a peak around the non-equilibrium chemical potential $\delta \mu$ at which temporal correlations in the external signal are reliably encoded inside the assembly. Together, the results of this section suggest that an analysis of the entropy production rate in such time-dependent non-equilibrium self-assembly processes can potentially be illuminating and reveal the requirements for encoding external temporal signals as spatial correlations in the assembly. 

\section{Conclusion}

In this paper, we have developed two frameworks to study the growing of an assembly under periodic drive. In the regime of high $\delta\mu$ and $T$, the mean-field equations for steady-state in our previous works~\cite{Nguyen2016,Nguyen2018} can be extended to predict the probability profile inside the assembly. A bi-partite Markov network has also been constructed that works in a wider regime. The results from these two approaches agree very well with those from FRM simulations.

We also demonstrate that a minimal driving force is required to reliably encode the external temporal signal as a spatial correlation in the assembly. Our work also suggests that accounting for the entropy production rate might help predict these minimum driving force conditions. Future work will explore the application of these ideas to higher dimensional materials with more complex interactions. 

The authors acknowledge support from the National Science Foundation under Grant No. DMR-1848306.
\bibliography{References02232018}
\onecolumngrid

\section*{\centering{\LARGE Appendix}}
\setcounter{figure}{0}
\setcounter{table}{0}
\setcounter{equation}{0}
\setcounter{page}{1}
\setcounter{section}{0}

\renewcommand{\theequation}{S\arabic{equation}} 
\renewcommand{\thepage}{S\arabic{page}} 
\renewcommand{\thesection}{S\arabic{section}}  
\renewcommand{\thetable}{S\arabic{table}}  
\renewcommand{\thefigure}{S\arabic{figure}}

\section{The Mean Field Equations}

The master equations correspond to the probabilities $P^t(n_i,n_j)$ that are used for our mean-field approach are: 
\begin{equation}
    \begin{split}
        &\frac{dP^t(n_i,n_j)}{dt}=-P^t(n_i,n_j)\Big[\sum_{n_k} P_{gen}(n_k|n_j)k_t^{add}({n_i,n_j\xrightarrow{}n_j,n_k})+\\
        &\sum_{n_k}P^t_{sys}(n_k|n_i)k_t^{rem}(n_k,n_i\xleftarrow{}n_i,n_j)\Big]+\\
        &\Big[P_{sys}^t(n_i|n_j)\sum_{n_k}P(n_j,n_k)k_t^{rem}({n_i,n_j\xleftarrow{}n_j,n_k})\Big]+\\
        &\Big[P_{gen}(n_j|n_i)\sum_{n_k}P(n_k,n_i)k_t^{add}({n_k,n_i\xrightarrow{}n_i,n_j})\Big]
    \label{eq:timeindependentequation}
    \end{split}
\end{equation}

\section{The Bi-Partite Master Equations}
According to Eq.~\ref{eq:addremoveparticletrans} in the main text, the addition of a new particle will only result in states in which the interface particle and its neighbor have the same timestamp. Thus, there are two kinds of master equations for the bi-partite Markov Network. The first one is for states in which the interface particle and its neighbor have the same timestamp:
\begin{equation}
    \begin{split}
        &\frac{dP^t(n_{i}t_x,n_{j}t_x)}{dt}=-P^t(n_{i}t_x,n_{j}t_x)\Big[\sum_{n_k} P_{gen}(n_k|n_j)k_t^{add}({n_{i}t_x,n_{j}t_x\xrightarrow{}n_jt_x,n_kt_x})\\
        &+\sum_{n_k,t_y}P^t_{sys}(n_kt_y|n_it_x)k_t^{rem}(n_kt_y,n_it_x\xleftarrow{}n_{i}t_x,n_{j}t_x)\Big]\\
        &+\Big[\sum_{n_k,t_y}P^t_{sys}(n_it_x|n_jt_y)P^t(n_jt_y,n_kt_x)k_t^{rem}({n_it_x,n_jt_x\xleftarrow{}n_jt_y,n_kt_x})\Big]\\
        &+\Big[\sum_{n_k,t_y}P_{gen}(n_j|n_i)P^t(n_kt_y,n_it_x)k_t^{add}({n_kt_y,n_it_x\xrightarrow{}n_it_x,n_jt_x}))\Big]\\
        &+\lambda P^t(n_{i}t_x,n_{j}t_{x-1})-\lambda P^t(n_{i}t_x,n_{j}t_{x})
    \label{eq:timedependentequation11}
    \end{split}
\end{equation}
The second one is for state in which the interface particle and its neighbor have different timestamp:
\begin{equation}
    \begin{split}
        &\frac{dP^t(n_{i}t_x,n_{j}t_z)}{dt}=-P^t(n_{i}t_x,n_{j}t_z)\Big[\sum_{n_k} P_{gen}(n_k|n_j)k_t^{add}({n_{i}t_x,n_{j}t_z\xrightarrow{}n_jt_z,n_kt_z})+\\
        &\sum_{n_k,t_y}P^t_{sys}(n_kt_y|n_it_x)k_t^{rem}(n_kt_y,n_it_z\xleftarrow{}n_{i}t_x,n_{j}t_z)\Big]\\&
        +\Big[\sum_{n_k,t_y}P^t_{sys}(n_it_x|n_jt_y)P^{t}(n_jt_y,n_kt_z)k_t^{rem}({n_it_x,n_jt_z\xleftarrow{}n_jt_y,n_kt_z})\Big]\\
        &+\lambda P^{t}(n_{i}t_x,n_{j}t_{z-1})-\lambda P^{t}(n_{i}t_x,n_{j}t_{z})
    \label{eq:timedependentequation12}
    \end{split}
\end{equation}
In Eq.~\ref{eq:timedependentequation12}, $t_x\neq t_z$. To obtain, Eq.~11 from the main-text, we consider the probability: $P(n_it_x,n_j)=\sum_{t_z}P^t(n_it_x,n_jt_z)$. The master equation of this probability constructed from Eq.~\ref{eq:timedependentequation11} and Eq.~\ref{eq:timedependentequation12} is:
\begin{equation}
\begin{split}
    &\frac{dP^t(n_{i}t_x,n_{j})}{dt}=-\Big[\sum_{n_k,t_z} P^t(n_{i}t_x,n_{j}t_z)P_{gen}(n_k|n_j)k_t^{add}({n_{i}t_x,n_{j}t_z\xrightarrow{}n_jt_z,n_kt_z})+\\
    &\sum_{n_k,t_y,t_z}P^t(n_{i}t_x,n_{j}t_z)P^t_{sys}(n_kt_y|n_it_x)k_t^{rem}(n_kt_y,n_it_z\xleftarrow{}n_{i}t_x,n_{j}t_z)\Big]\\
    &+\Big[\sum_{n_k,t_y,t_z}P^t_{sys}(n_it_x|n_jt_y)P^{t}(n_jt_y,n_kt_z)k_t^{rem}({n_it_x,n_jt_z\xleftarrow{}n_jt_y,n_kt_z})\Big]\\
    &+\Big[\sum_{n_k,t_y}P_{gen}(n_j|n_i)P^t(n_kt_y,n_it_x)k_t^{add}({n_kt_y,n_it_x\xrightarrow{}n_it_x,n_jt_x})\Big]
    \end{split}
    \label{eq:currenttimerelation}
\end{equation}
When the system at steady state, we can set the left side of Eq.~\ref{eq:currenttimerelation} to 0 and rewrite it into:
\begin{equation}
    \begin{split}
    &\sum_{n_k,t_y,t_z}\Big[ P^{ss}(n_{i}t_x,n_{j}t_z)P_{gen}(n_k|n_j)k_t^{add}({n_{i}t_x,n_{j}t_z\xrightarrow{}n_jt_z,n_kt_z})-\\
    &P^{ss}_{sys}(n_it_x|n_jt_y)P^{t}(n_jt_y,n_kt_z)k_t^{rem}({n_it_x,n_jt_z\xleftarrow{}n_jt_y,n_kt_z})\Big]=\\
    &\sum_{n_k,t_y,t_z}\Big[P_{gen}(n_j|n_i)P^{ss}(n_kt_y,n_it_x)k_t^{add}({n_kt_y,n_it_x\xrightarrow{}n_it_x,n_jt_x})-\\
    &P^{ss}(n_{i}t_x,n_{j}t_z)P^{ss}_{sys}(n_kt_y|n_it_x)k_t^{rem}(n_kt_y,n_it_z\xleftarrow{}n_{i}t_x,n_{j}t_z)\Big]
    \label{eq:steadystatetime}
    \end{split}
\end{equation}
Next, we make an assumption that the rates of the process only depends on the two further most layers of the assembly. Specifically, the rate $k_t^{add}({n_{i}t_x,n_{j}t_y\xrightarrow{}n_jt_y,n_kt_y})$ only depends on $(n_jt_y,n_kt_y)$; and the rate $k_t^{rem}({n_it_x,n_jt_z\xleftarrow{}n_jt_y,n_kt_z})$ depends on $(n_jt_y,n_kt_z)$. If this is not the case for our rates, we can expand the element $n_i$ to include more layers until the rates only depend on the two furthest elements.
We then rewrite the left side of Eq.~\ref{eq:steadystatetime} into:
\begin{equation}
    \begin{split}
    &\sum_{t_y}P^{ss}_{sys}(n_it_x|n_jt_y)\Big[\sum_{n_k,t_z}\Big[ P^{ss}(n_jt_y)P_{gen}(n_k|n_j)k_t^{add}({n_{i}t_x,n_{j}t_y\xrightarrow{}n_jt_y,n_kt_y})-\\
    &P^{t}(n_jt_y,n_kt_z)k_t^{rem}({n_it_x,n_jt_z\xleftarrow{}n_jt_y,n_kt_z})\Big]\Big]=\\
    &\sum_{t_y}P^{ss}_{sys}(n_it_x|n_jt_y)\Big[\sum_{n_k,n_i^{'},t_z,t_x^{'}}\Big[ P^{ss}(n_i^{'}t_x^{'},n_jt_y)P_{gen}(n_k|n_j)k_t^{add}({n_{i}^{'}t_x^{'},n_{j}t_y\xrightarrow{}n_jt_y,n_kt_y})-\\
    &P^{t}(n_jt_y,n_kt_z)P(n_i^{'}t_x^{'}|n_jt_z)k_t^{rem}({n_i^{'}t_x^{'},n_jt_z\xleftarrow{}n_jt_y,n_kt_z})\Big]\Big]
    \label{eq:firstpartsteady}
    \end{split}
\end{equation}
For the sake of notation, let's define:
\begin{equation}
\begin{split}
    V(n_jt_y)=&\sum_{n_k,n_i^{'},t_z,t_x^{'}}\Big[ P^{ss}(n_i^{'}t_x^{'},n_jt_y)P_{gen}(n_k|n_j)k_t^{add}({n_{i}^{'}t_x^{'},n_{j}t_y\xrightarrow{}n_jt_y,n_kt_y})-\\
    &P^{t}(n_jt_y,n_kt_z)P(n_i^{'}t_x^{'}|n_jt_z)k_t^{rem}({n_i^{'}t_x^{'},n_jt_z\xleftarrow{}n_jt_y,n_kt_z})\Big]
    \label{eq:vfull}
\end{split}
\end{equation}

\begin{equation}
\begin{split}
    V(n_it_x,n_j)=&\sum_{n_k,t_y,t_z}\Big[P^{ss}(n_kt_y,n_it_x)P_{gen}(n_j|n_i)k_t^{add}({n_kt_y,n_it_x\xrightarrow{}n_it_x,n_jt_x})-\\
    &P^{ss}(n_{i}t_x,n_{j}t_z)P^{ss}_{sys}(n_kt_y|n_it_x)k_t^{rem}(n_kt_y,n_it_z\xleftarrow{}n_{i}t_x,n_{j}t_z)\Big]
    \label{eq:vhalf}
\end{split}
\end{equation}
Plug Eqs.~\ref{eq:vhalf}, \ref{eq:vfull}, \ref{eq:firstpartsteady} into Eq.~\ref{eq:steadystatetime} and we have:
\begin{equation}
    V(n_it_x,n_j)=\sum_{t_y}P^{ss}_{sys}(n_it_x|n_jt_y)V(n_jt_y)
    \label{eq:velocityrelationship}
\end{equation}
Summing all over $n_j$ and we have:

\begin{equation}
    V(n_it_x)=\sum_{n_j,t_y}P^{ss}_{sys}(n_it_x|n_jt_y)V(n_jt_y)
    \label{eq:velocityrelationfull}
\end{equation}
Eq.~\ref{eq:velocityrelationfull} suggests that: $\bf{V}$ is an eigenvector of the conditional probability matrix $\bf{P^{ss}_{sys}}$. Another eigenvector of the conditional probability matrix is $\bf{P^{ss}_{inside}}$ since:
\begin{equation}
    P^{ss}_{inside}(n_it_x)=\sum_{n_j,t_y}P^{ss}_{sys}(n_it_x|n_jt_y)P^{ss}_{inside}(n_jt_y)
\end{equation}
This indicates that $P^{ss}_{inside}(n_it_x)\propto V(n_it_x)$. Using the normalization of probability, we then have:
\begin{equation}
    P^{ss}_{inside}(n_it_x)=\frac{V(n_it_x)}{\sum_{n_i,t_x}V(n_it_x)}
\end{equation}
\section{The Coarse Grain Master Equations and Its Entropy Production}

As demonstrated in the main-text, the entropy production of the bi-partite Markov process is always higher than the entropy production of the mean-field model. To convene these two approaches and their entropy productions, we have to look at a coarse grain probability from the bi-partite Markov model: 
\begin{equation}
    \begin{split}
        P^{t}(n_i,n_jt_y)=\sum_{t_x}P^{t}(n_it_x,n_jt_y)
    \end{split}
\end{equation}
In effect, we have merge the time partitions in the bi-partite Markov process in the coarse grain probabilities. 
The master equations for the coarse grain probability, $P^{t}(n_{i},n_{j}t_y)$ is:
\begin{equation}
    \begin{split}
        &\frac{dP^{t}(n_{i},n_{j}t_y)}{dt}=\\
        &-\sum_{t_x,t_z,n_k}P^t(n_{i}t_x,n_{j}t_y)\Big[ P_{gen}(n_kt_y|n_jt_y)k^{add}({n_{i}t_x,n_{j}t_y\xrightarrow{}n_jt_y,n_kt_y})+\\
        &P^t_{sys}(n_kt_z|n_it_x)k^{rem}(n_kt_z,n_it_y\xleftarrow{}n_{i}t_x,n_{j}t_y)\Big]+\\
        &\Big[\sum_{n_k,t_x,t_z}P^t_{sys}(n_it_x|n_jt_z)P^{t}(n_jt_z,n_kt_y)k^{rem}({n_it_x,n_jt_y\xleftarrow{}n_jt_z,n_kt_y})\Big]+\\
        &\Big[\sum_{n_k,t_x}P_{gen}(n_jt_y|n_it_y)P^{t}(n_kt_x,n_it_y)k^{add}({n_kt_x,n_it_y\xrightarrow{}n_it_y,n_jt_y}))\Big]\\
        &+\lambda P^{t}(n_i,n_{j}t_{y-1})-\lambda P^{t}(n_{i},n_{j}t_{y})\\
        &=-P^{t}(n_{i},n_{j}t_y)\Bigg(\sum_{t_x,,t_z,n_k}P^t_{sys}(t_x|n_{i},n_{j}t_y)\\
        &\Big[ P_{gen}(n_kt_y|n_jt_y)k^{add}({n_{i}t_x,n_{j}t_y\xrightarrow{}n_jt_y,n_kt_y})+\\
        &P^t_{sys}(n_kt_z|n_it_x)k^{rem}(n_kt_z,n_it_y\xleftarrow{}n_{i}t_x,n_{j}t_y)\Big]\Bigg)+\\
        &\sum_{n_k}\Bigg(P^{t}(n_{j},n_{k}t_y)\Big[\sum_{t_x,t_z}P^t_{sys}(n_it_x|n_jt_z)P^t_{sys}(t_z|n_j,n_kt_y)k^{rem}({n_it_x,n_jt_y\xleftarrow{}n_jt_z,n_kt_y})\Big]\\
        &+P^{t}(n_{k},n_{i}t_y)\Big[\sum_{t_x}P_{gen}(n_jt_y|n_it_y)P^t_{sys}(t_x|n_k,n_it_y)k^{add}({n_kt_x,n_it_y\xrightarrow{}n_it_y,n_jt_y})\Big]\Bigg)\\
        &+\lambda P^{t}(n_i,n_{j}t_{y-1})-\lambda P^{t}(n_{i},n_{j}t_{y})
    \label{eq:coarsegrainmastereq}
    \end{split}
\end{equation}
Unlike the master equations for the full probabilities bi-partite Markov process, the master equations for the coarse grain probabilities (Eq.~\ref{eq:coarsegrainmastereq}) cannot be self-consistently solved at a steady state because it still depends on the probability $P^{t}(n_it_x,n_jt_y)$. So to get the coarse grain probabilities, we have to solve the master equations of the full probabilities first. 
The entropy production corresponds to the coarse grain probabilities is:
\begin{equation}
\begin{split}
    &\dot{S}_{reduced}=\sum_{t_y,n_k,n_i,n_j}\\
    &\Big(P^{t}(n_{k},n_{i}t_y)\big[\sum_{t_x}P_{gen}(n_jt_y|n_it_y)P^t_{sys}(t_x|n_k,n_it_y)k^{add}({n_kt_x,n_it_y\xrightarrow{}n_it_y,n_jt_y})\big]\\
    &-P^{t}(n_{i},n_{j}t_y)\big[\sum_{t_x,t_z}P^t_{sys}(n_kt_x|n_it_z)P^t_{sys}(t_z|n_i,n_jt_y)k^{rem}({n_kt_x,n_it_y\xleftarrow{}n_it_z,n_jt_y})\big]\Big)\\
    &\ln{\frac{P^{t}(n_{k},n_{i}t_y)\big[\sum_{t_x}P_{gen}(n_jt_y|n_it_y)P^t_{sys}(t_x|n_k,n_it_y)k^{add}({n_kt_x,n_it_y\xrightarrow{}n_it_y,n_jt_y})\big]}{P^{t}(n_{i},n_{j}t_y)\big[\sum_{t_x,t_z}P^t_{sys}(n_kt_x|n_it_z)P^t_{sys}(t_z|n_i,n_jt_y)k^{rem}({n_kt_x,n_it_y\xleftarrow{}n_it_z,n_jt_y})\big]}}
\label{eq:entropyproductioncoasegrain}
\end{split}
\end{equation}
\begin{figure}
    \centering
    \includegraphics[width=1\linewidth,angle=270,scale=0.75]{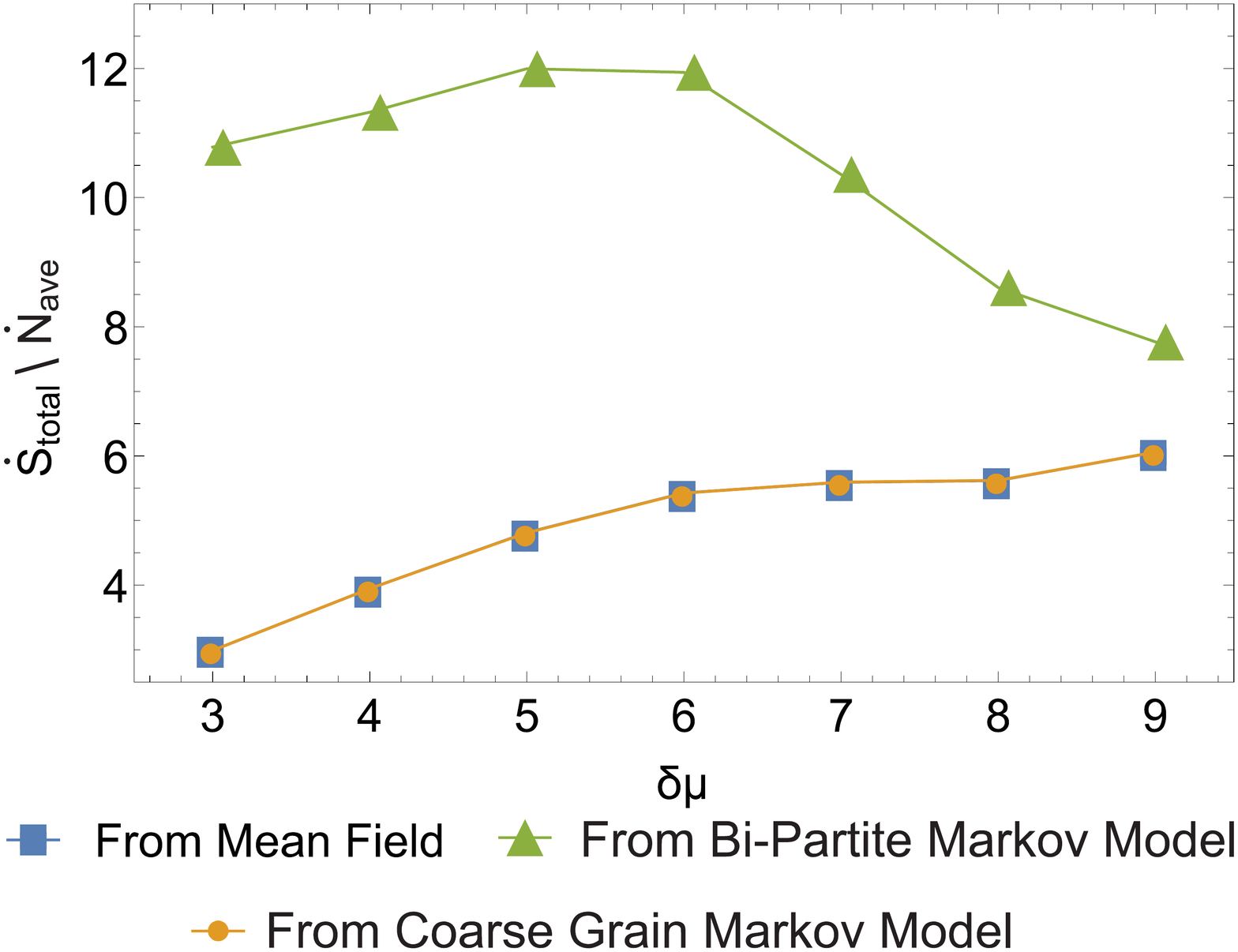}
    \caption{Entropy production of the one dimension growth assembly with rates affected by a time-varying external field h(t) with $J = 4, h = 0.2, T = 1000$. $\dot{N}_{\rm{ave}}$ is the average growth rate of assemblies over the period T: $\dot{N}_{\rm{ave}}=\frac{1}{T}\int_0^T\dot{N}(t)dt$. The blue curve is computed using the mean-field entropy production, Eq.~\ref{eq:entropyprodmeanfield}. The green curve is computed using the full bi-partite Markov Process entropy production, Eq.~\ref{eq:totalentropyproductionplusobserve}. The orange curve is computed using the entropy production of the coarse grain Markov Process, Eq.~S15.
    }
    \label{fig:EntropyProductionTimeDepCoarseGrain}
\end{figure}
As shown in Fig.~S1, the entropy production of the coarse grain model, Eq.~S15, is identical to one from the mean field model, Eq.~\ref{eq:entropyprodmeanfield}.
\end{document}